\documentclass[floatfix,prd,epsfig,nofootinbib,superscriptaddress,onecolumn,
amssymb]{revtex4}

\usepackage{graphicx}
\usepackage{color}

\begin{document}

\title{Measuring the Neutrino Mass Hierarchy with Atmospheric Neutrinos}

\author{D.F. Cowen}
\email{cowen@phys.psu.edu}
\affiliation{Department of Physics, Pennsylvania State University, University Park, PA}

\author{T. DeYoung}
\email{deyoung@pa.msu.edu}
\affiliation{Department of Physics and Astronomy, Michigan State
  University, East Lansing, MI}

\author{D. Grant}
\email{drg@ualberta.ca}
\affiliation{Department of Physics, University of Alberta, Edmonton, Alberta}

\author{D. A. Dwyer}
 \email{DADwyer@lbl.gov}
\affiliation{Lawrence Berkeley National Laboratory, Berkeley, CA}

\author{S. R. Klein}\footnote{Corresponding Author}
 \email{srklein@lbl.gov}
\affiliation{Lawrence Berkeley National Laboratory, Berkeley, CA}
\affiliation{the University of California, Berkeley, CA}

\author{K. B. Luk}
\email{K_Luk@lbl.gov}
\affiliation{Lawrence Berkeley National Laboratory, Berkeley, CA}
\affiliation{the University of California, Berkeley, CA}

\author{D.R. Williams}
 \email{drwilliams3@ua.edu}
\affiliation{Department of Physics and Astronomy, University of Alabama, Tuscaloosa, AL}

\collaboration{for the IceCube/PINGU Collaboration}\homepage{http://icecube.wisc.edu/collaboration/authors/pingu}

\date{Sept. 19, 2014}

\begin{abstract}

\end{abstract}

\maketitle


\def\onbb{$0\nu\beta\beta$}

The neutrino mass hierarchy is one of the key remaining unknowns in
the neutrino sector, with important implications in a number of
nuclear physics problems, including neutrinoless double beta decay
(\onbb) and the physics of supernova explosions.  \onbb\ in particular
is a key focus of neutrino research in nuclear
physics~\cite{0nbbRef}.

In \onbb, the relationship between the effective mass for neutrinoless
double beta decay and the mass of the lightest neutrino depends on
whether the mass hierarchy is normal or inverted.  If the mass
hierarchy is inverted, then there is a minimum effective mass which
could be reached by envisioned next-generation neutrinoless double
beta decay experiments.  If there were an independent measurement of
the mass hierarchy, an experiment that reached this limit could
conclusively state that neutrinos are not Majorana particles.  If the
mass hierarchy is normal or unknown, then no such statement is
possible.  Experiments could observe \onbb, but, in the absence of an
observation, the nature of neutrinos would remain uncertain.

\begin{figure}
     \includegraphics[width=\columnwidth, angle=0]{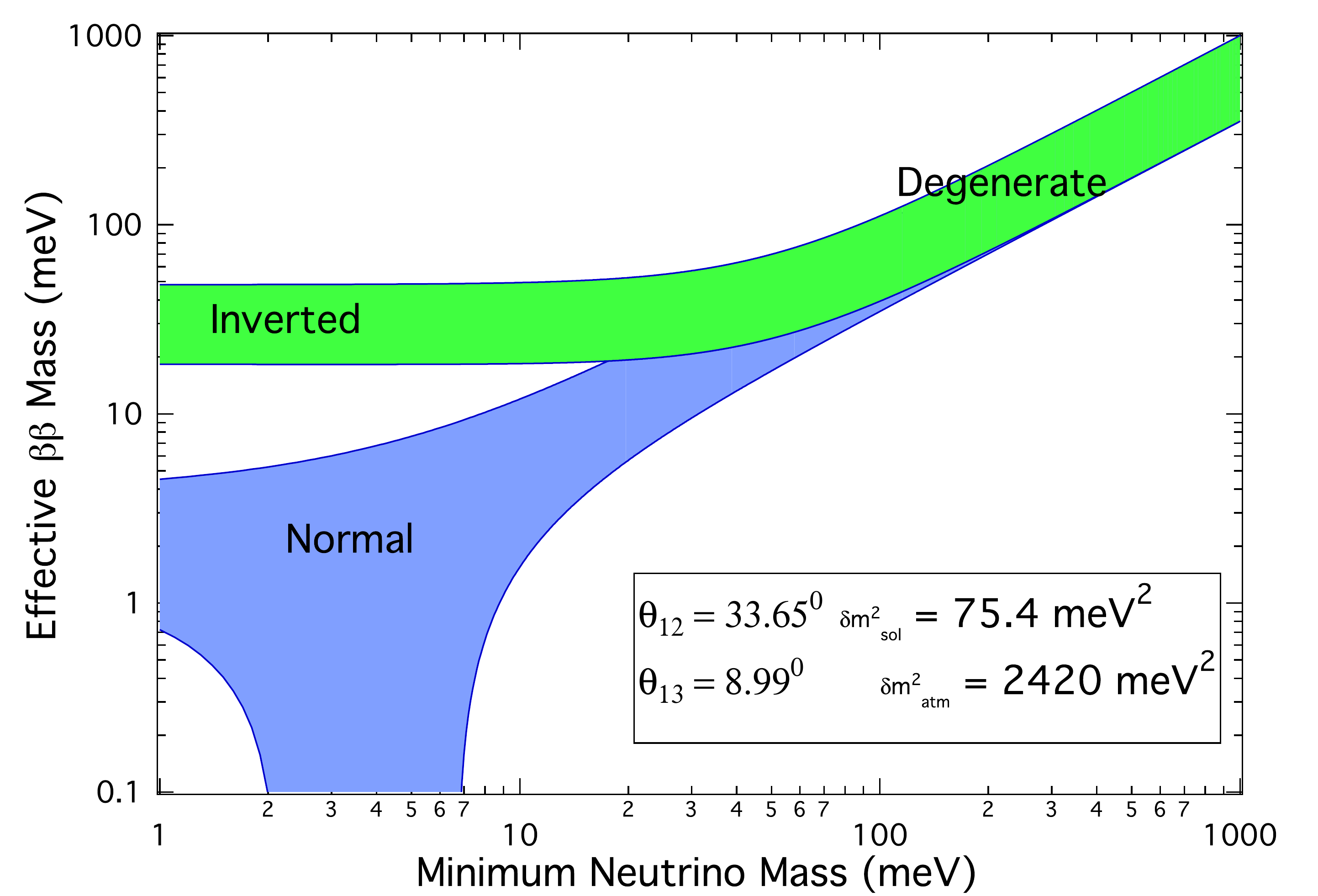}
     \caption{The effective neutrino Majorana mass $m_{\beta\beta}$ as
       a function of the lightest neutrino mass, $m_{\rm light}$. The
       blue (green) band corresponds to the normal (inverted) ordering,
       respectively, in which case $m_{\rm light}$ is equal to $m_1$
       ($m_3$).   The widths of the bands come from uncertainties in the
       other neutrino properties. From Ref.~\cite{Elliott}.}
  \label{fig:0nBB}
\end{figure}

Directly measuring the neutrino mass hierarchy requires
high-resolution measurements of neutrino oscillations.  There are a
number of proposed methods to do this~\cite{Cahn:2013taa}. The least
costly and possibly fastest approach is to use atmospheric neutrinos.
Three groups are proposing this: PINGU (Precision IceCube Next
Generation Upgrade) in the Antarctic ice cap,  ORCA (Oscillation Research
with Cosmics in the Abyss) in the Mediterranean Sea~\cite{Katz:2014tta},
and the India-based Neutrino Observatory \cite{Devi:2014yaa}.
Here, we focus on PINGU, which has a large U. S. participation.  

PINGU will permit a
determination of the mass hierarchy, independent of the CP violation
parameter, at relatively modest expense, using a
well-understood technique with minimal risk, on a short time scale.  It will
leverage the knowledge gained in designing, deploying and operating
IceCube and its in-fill array DeepCore.  IceCube and DeepCore have
been continuously taking high-quality physics data since early 2011.
By deploying PINGU within and around existing IceCube and DeepCore
digital optical modules (DOMs), a multi-megaton fiducial volume of ice
would be instrumented with sufficient photocathode density to yield a
neutrino energy threshold of a few GeV.  The scale of PINGU would
permit measurement of the oscillations of atmospheric neutrinos over a
range of energies and a variety of baselines (up to the diameter of
the Earth) with sufficient precision that hierarchy-dependent
distortions of the oscillations due to the presence of matter could be
observed.   PINGU will also provide a precise measurement of
$\theta_{23}$. 

PINGU construction and technology would be similar to that used in
IceCube, with large photomultiplier tubes and readout electronics
encased in pressure vessels embedded in the Antarctic ice cap below
the US Amundsen-Scott South Pole Station.  The 2850~m thick, very
transparent South Pole ice cap would serve simultaneously as
neutrino target, Cherenkov medium, and detector support structure.


A likely construction scenario places PINGU under the umbrella of an
expanded IceCube-based facility at the South Pole.  PINGU would be
constructed first, followed by an extension focused on high energy
astrophysical neutrinos, obtaining economies of scale through the use
of common hardware and installation techniques.  The PINGU share of
the facility cost is roughly \$55M (US cost, including contingency)
plus \$25M (foreign contribution) for a total of \$80M.  Detector
construction can be completed five years after funding starts, or as
early as 2020. The determination of the mass hierarchy at the
$3\sigma$ level would be possible with about 3~years of data.

A determination of the neutrino mass hierarchy would contribute to
advances in a number of other areas in the nuclear physics purview.
Knowing the neutrino mass hierarchy is also important for
understanding how supernovae explode; neutrinos interact collectively
with the matter in supernovae, and the character of these interactions
depends on the hierarchy.  These differences are important in
modelling supernovae~\cite{Chiu:2013dya} and understanding
heavy-element production in the universe, and they also have
observational consequences~\cite{Duan:2007bt}.  An independent
determination of the neutrino mass hierarchy would allow future
observations of neutrinos from supernovae to be used to much better
pin down other aspects of the supernova explosion process.

PINGU will also become one of a handful of active supernova neutrino
detectors.  Its multi-megaton fiducial volume gives it the ability to
observe galactic supernova with unprecedented (millisecond) time
resolution~\cite{Abbasi:2011ss}, and it will have a phototube density
high enough to determine both the integrated neutrino luminosity and
the neutrino energy spectrum on short time scales~\cite{Aartsen:2014oha}.

In conclusion, PINGU offers an extremely cost effective way to provide
answers to the key (and still relevant) question posed in the 2007
Nuclear Science Long Range Plan~\cite{LRP}, ``What is the nature of
the neutrinos, what are their masses, and how have they shaped the
evolution of the universe?'' PINGU will use atmospheric neutrinos to
determine the neutrino mass hierarchy, with a direct impact on the
interpretation of \onbb\ measurements, the modelling and understanding
of supernova explosions, and the detector will serve as a premier
supernova neutrino detector in its own right.

We acknowledge support from the U.S. National Science
Foundation-Office of Polar Programs and the U.S. National Science
Foundation-Physics Division.  This Position Paper is based upon work
supported in part by the U.S. Department of Energy, Office of Science,
Office of Nuclear Physics, under contract number DE-AC-76SF00098.

\end{document}